\documentclass[a4paper]{jpconf}
\usepackage{textcomp}
\usepackage{amsfonts} 
\usepackage{amssymb} 
\usepackage{amsmath}
\usepackage{slashed}
\usepackage{graphicx}
\usepackage{comment}
\usepackage{color}
\usepackage{url}
\usepackage{caption}
\graphicspath{{./fig/}}

\renewcommand{\Re}{\mathrm{Re}}
\renewcommand{\Im}{\mathrm{Im}}

\providecommand{\MeV}{\,\mathrm{MeV}}

\newcommand{\BK}{\hat{B}_{K}}

\newcommand{\eps}{\varepsilon}

\newcommand{\epsK}{\varepsilon_{K}}

\begin{document}
\title{Current status of $\epsK$ in lattice QCD}

\author{Weonjong Lee}

\address{Director, Lattice Gauge Theory Research Center, CTP, and FPRD,\\
  Department of Physics and Astronomy, \\
  Seoul National University,
  Seoul 08826, South Korea}

\ead{wlee@snu.ac.kr}

\begin{abstract}
We present the current status of $\epsK$ evaluated directly from the
standard model using lattice QCD inputs.
The lattice QCD inputs include $\BK$, $\xi_0$, $\xi_2$, $|V_{us}|$,
$m_c(m_c)$, and $|V_{cb}|$.
Recently, FLAG has updated $\BK$, exclusive $|V_{cb}|$ has been updated 
with new lattice data in the $\bar{B}\to D\ell\bar{\nu}$ decay mode, and
RBC-UKQCD has updated $\xi_0$ and $\xi_2$.
We find that the standard model evaluation of $\epsK$ with exclusive
$|V_{cb}|$ (lattice QCD inputs) is $3.2\sigma$ lower than the
experimental value, while that with inclusive $|V_{cb}|$ (heavy quark 
expansion) shows no tension.
\end{abstract}

\section{Introduction}
Since 2012, we have been monitoring $\epsK$, the indirect CP violation
parameter in neutral kaons using lattice QCD inputs.
The parameter $\epsK$ is, in particular, very attractive to the particle
physics community, since it is very precisely measured in
experiment, and it provides a direct probe of CP violation in
the standard model and in physics models 
beyond the standard model (BSM).
In this paper, we present results of $\epsK$ evaluated directly from
the standard model with lattice QCD inputs and compare them with the
experimental value. 
This paper is an update of our previous paper \cite{ Bailey:2015tba,
Bailey:2015frw}.

\section{Input parameters}
In the standard model, the indirect CP violation parameter of the
neutral kaon system $\epsK$ can be expressed as follows,
\begin{align}
  \epsK
  & \equiv \frac{\mathcal{A}(K_L \to \pi\pi(I=0))}
              {\mathcal{A}(K_S \to \pi\pi(I=0))}
  \nonumber \\
  & =  e^{i\theta} \sqrt{2}\sin{\theta}
  \Big( C_{\eps} X_\text{SD} \hat{B}_{K} 
  + \frac{ \xi_{0} }{ \sqrt{2} } + \xi_\text{LD} \Big)
   + \mathcal{O}(\omega\eps^\prime)
   + \mathcal{O}(\xi_0 \Gamma_2/\Gamma_1) \,.
  \label{eq:epsK_def}
\end{align}
Here, the short distance contribution proportional to $\BK$ occupies 
about 105\% of $\epsK$, the long distance effect from the
absorptive part $\xi_0$ gives about $-5\%$ correction, and
the long distance effect from the dispersive part $\xi_\text{LD}$ 
gives about $\pm$1.6\% correction.
The details on $C_{\eps}$, $X_\text{SD}$, $\xi_0$, and
$\xi_\text{LD}$  are described in Ref.~\cite{Bailey:2015tba}.
In order to determine $\epsK$ directly from the standard model, we
need 18 input parameters, and 6 of them can, in principle, be obtained
from lattice QCD: $V_{us}$, $V_{cb}$, $\BK$, $\xi_0$, $\xi_\text{LD}$,
and $m_c(m_c)$.
Here, we address recent progress on determining those input
parameters.

\subsection{$|V_{cb}|$}
\begin{figure}[ht]
\begin{minipage}[t]{0.6\textwidth}
\hspace{-3pc}
\includegraphics[width=1.0\linewidth]{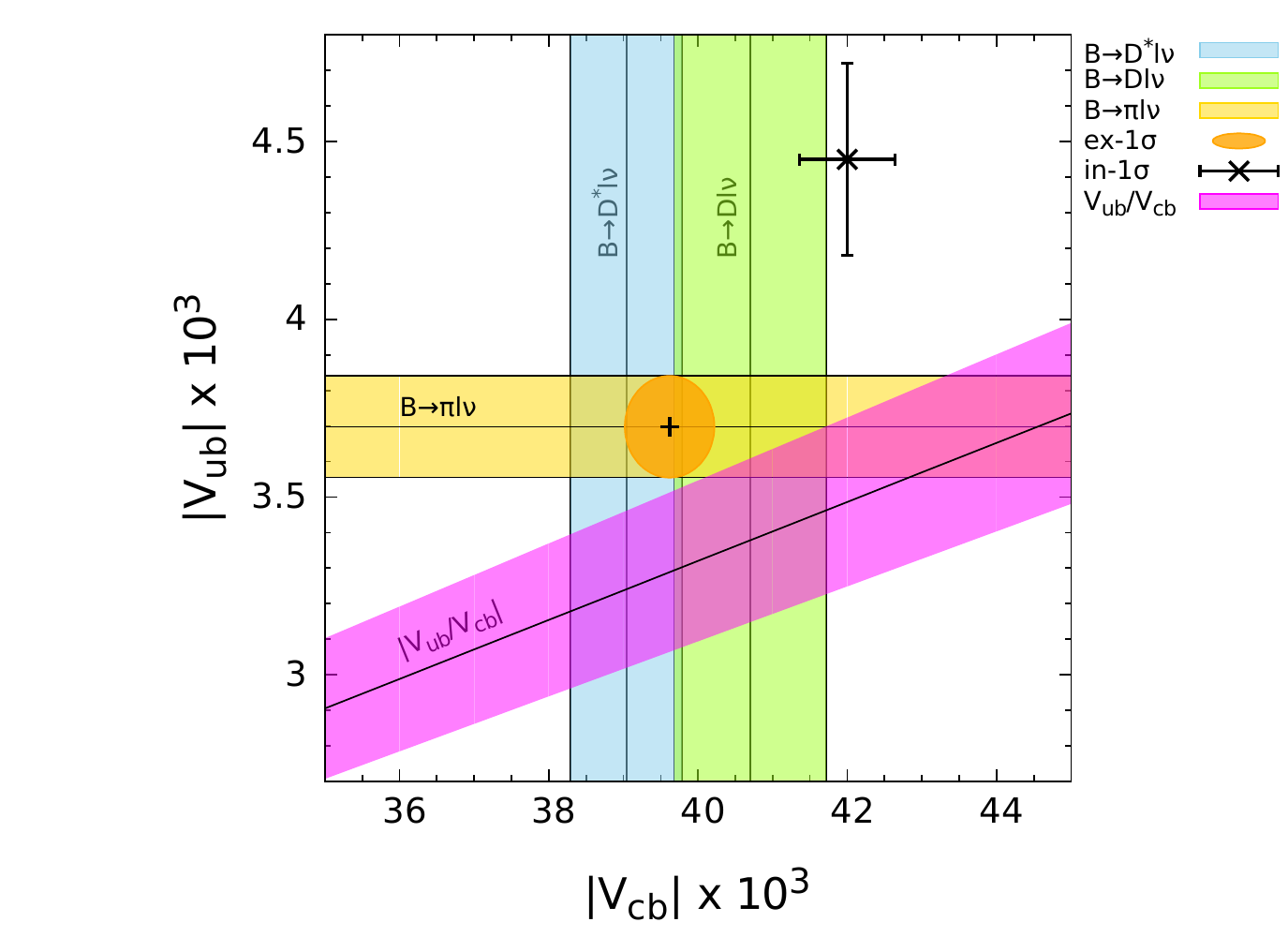}
\end{minipage}
\hspace{-2.5pc}
\begin{minipage}[b]{0.45\textwidth}
\caption{$|V_{cb}|$ versus $|V_{ub}|$. The sky-blue band represents
  $|V_{cb}|$ determined from the $\bar{B}\to D^* \ell \bar\nu$ decay
  mode. The yellow-green band represents $|V_{cb}|$ determined from
  the $\bar{B}\to D\ell\bar\nu$ decay mode. The yellow band represents
  $|V_{ub}|$ determined from the $\bar{B} \to \pi\ell\bar\nu$ decay mode.
  The magenta band represents $|V_{ub}/V_{cb}|$ determined from the
  LHCb data of the $\Lambda_b \to \Lambda_c \ell\bar\nu$ and
  $\Lambda_b \to p \ell\bar\nu$ decay modes. The orange circle
  represents the combined results for exclusive $|V_{cb}|$ and
  $|V_{ub}|$ from the $B$ meson decays within $1.0\sigma$. 
  The black cross (\textbf{\texttimes})
  represents the inclusive $|V_{cb}|$ and $|V_{ub}|$ from the heavy
  quark expansion. The details are given in Tables \ref{tab:Vcb} and
  \ref{tab:Vub}.}
\label{fig:Vcb-Vub}
\end{minipage}
\end{figure}

%
%
%
\begin{table}[htbp]
  \begin{minipage}[b]{0.5\linewidth}
  \caption{Results for $|V_{cb}|$}
  \label{tab:Vcb}
  \begin{center}
  \begin{tabular}{lll}
    \br
    Decay mode & $|V_{cb}|$ & Ref. 
    \\ \mr
    $\bar{B}\to D^*\ell\bar{\nu}$ & $39.04(49)(53)(19)$ & 
    \cite{Bailey2014:PhysRevD.89.114504}
    \\ 
    $\bar{B}\to D\ell\bar{\nu}$ & $40.7(10)(2)$ & \cite{DeTar:2015orc}
    \\ 
    ex-combined & $39.62(60)$ & this paper
    \\ 
    $\bar{B}\to X_c\ell\bar{\nu}$ & 42.00(64) & \cite{Gambino:2016jkc}
    \\ \br
  \end{tabular} 
  \end{center}
  \end{minipage}
  \hfill
  \begin{minipage}[b]{0.5\linewidth}
  \caption{Results for $|V_{ub}|$}
  \label{tab:Vub}
  \begin{center}
  \begin{tabular}{lll}
    \br
    Decay mode & $|V_{ub}|$ & Ref. 
    \\ \mr
    $\bar{B}\to \pi\ell\bar{\nu}$ & 3.72(16) & \cite{Lattice:2015tia}
    \\
    $\bar{B}\to \pi\ell\bar{\nu}$ & 3.61(32) & \cite{Flynn:2015mha}        
    \\
    ex-combined & $3.70(14)$ & this paper        
    \\
    $\bar{B}\to X_u\ell\bar{\nu}$ & $4.45(16)(22)$ & \cite{Amhis:2014hma}
    \\ \br
  \end{tabular} 
  \end{center}
  \end{minipage}
\end{table}
Recent results for $|V_{cb}|$ and $|V_{ub}|$ are presented in
Tables \ref{tab:Vcb} and \ref{tab:Vub}.
Recently, DeTar has collected the results of $\bar{B}\to
D\ell\bar\nu$ decay at non-zero recoil from both lattice QCD \cite{
  Lattice:2015rga, Na:2015kha} and experiments of Babar \cite{
  Aubert:2008yv} and Belle \cite{ Glattauer:2015yag}, and has made a
combined fit of all the data simultaneously to determine $|V_{cb}|$
\cite{ DeTar:2015orc}.
We have obtained the ``ex-combined'' result in Table \ref{tab:Vcb}
by taking a weighted average of the $V_{cb}$ results from the $\bar{B}\to
D^*\ell\bar\nu$ and $\bar{B}\to D\ell\bar\nu$ decay channels.
Similarly, we have obtained the ``ex-combined'' result in Table
\ref{tab:Vub} by taking a weighted average of the two $V_{ub}$
results from $\bar{B}\to \pi \ell\bar\nu$ decay.
In Fig.~\ref{fig:Vcb-Vub}, we show all the results in various
colors.\footnote{ The plot is based on that by Andreas Kronfeld in
  Ref.~\cite{ DeTar:2015orc}.}
The inclusive results are about $3\sigma$ away from those from
exclusive $B$ meson decays respectively as well as from the LHCb
results for $|V_{ub}/V_{cb}|$, which corresponds to the magenta band
in Fig.~\ref{fig:Vcb-Vub}.

\subsection{$\xi_0$ and $\xi_\mathrm{LD}$}

%
%
\begin{table}[htbp]
  \begin{minipage}[c]{0.40\linewidth}
    \caption{$\xi_0$ and $\xi_\mathrm{LD}$.}
    \label{tab:in-LD}
    \begin{center}
    \small
    \begin{tabular}{llll}
      \br
      Input & Method & Value & Ref. 
      \\ \mr
      $\xi_0$ & indirect & $-1.63(19) \times 10^{-4}$ 
      & \cite{Blum:2015ywa} 
      \\
      $\xi_0$ & direct  & $-0.57(49) \times 10^{-4}$ 
      & \cite{Bai:2015nea}
      \\ \mr
      $\xi_\mathrm{LD}$ & --- & $(0 \pm 1.6)\,\%$ 
      & \cite{Christ2012:PhysRevD.88.014508} 
      \\ \br
    \end{tabular}
    \end{center}
    \vspace{3mm}
    \normalsize
    \captionof{table}{$\delta_0$}
    \label{tab:d0}
    \begin{center}
    \small
    \begin{tabular}{lll}
      \br
      Collaboration & $\delta_0$ & Ref.
      \\ \mr
      RBC-UK-2016  & $23.8(49)(12){}^{\circ}$ & \cite{Bai:2015nea}
      \\
      KPY-2011 & 39.1(6)${}^{\circ}$ & \cite{GarciaMartin:2011cn}
      \\
      CGL-2001 & 39.2(15)${}^{\circ}$ & \cite{Colangelo:2001df,Colangelo2016:MITP}
      \\ \br
    \end{tabular}
    \end{center}
  \end{minipage}
  \hfill
  \begin{minipage}[c]{0.54\linewidth}
    \begin{center}
    \includegraphics[width=\textwidth]{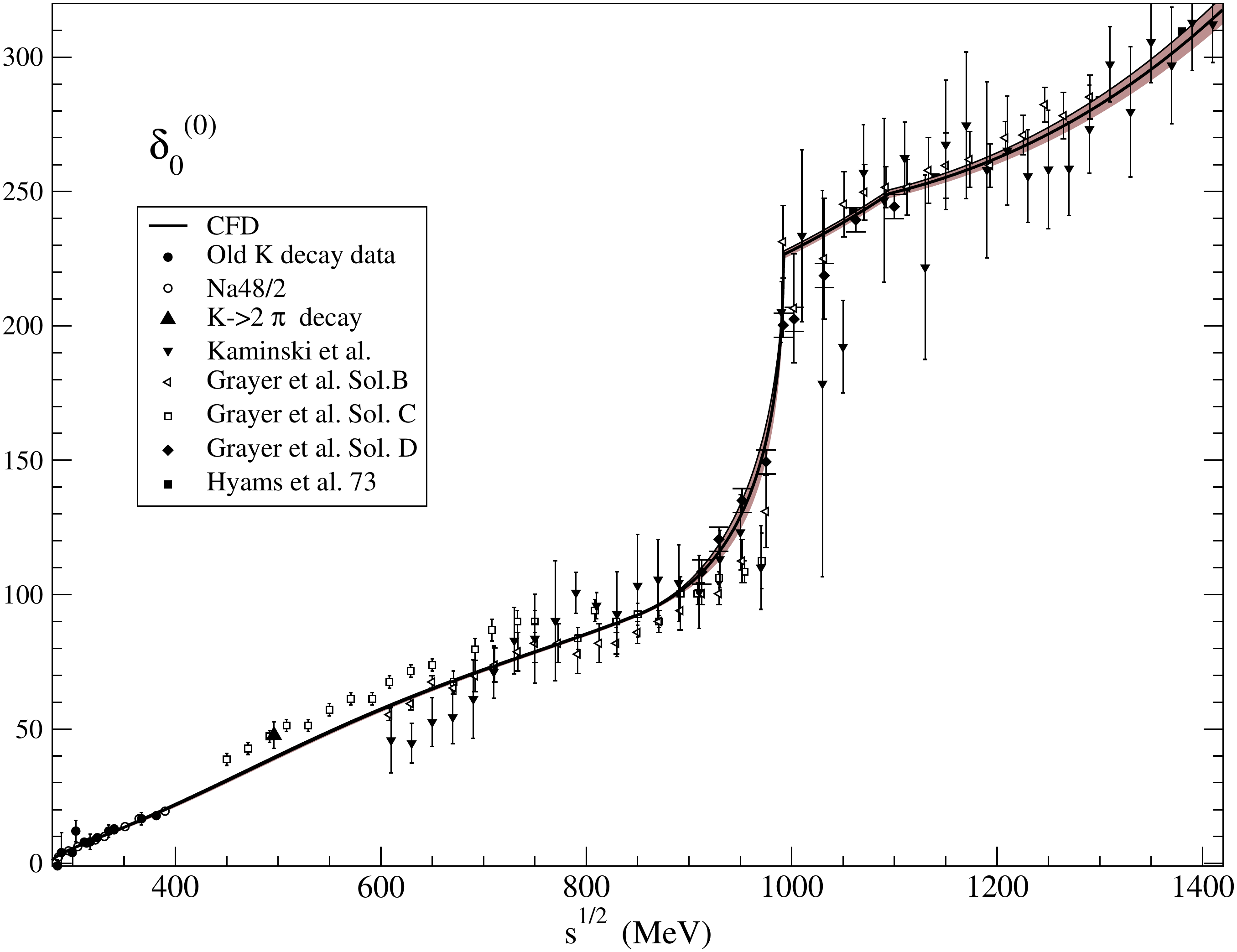}
    \end{center}
    \captionof{figure}{$\delta_0$ from KPY-2011.}
    \label{fig:d0-KPY-2011}
  \end{minipage}
\end{table}
There are two independent methods to determine $\xi_0$ in lattice QCD:
the indirect and direct methods.
In the indirect method, we determine $\xi_0$ from
the experimental values of $\Re(\eps'/\eps)$, $\epsK$, $\omega$ and 
the lattice QCD input $\xi_2$.
They are related to one another as follows,
\begin{align}
&\xi_0  = \frac{\Im A_0}{\Re A_0}, \qquad
\xi_2 = \frac{\Im A_2}{\Re A_2}, \qquad
\Re \left(\frac{\eps'}{\eps} \right) =
\frac{\omega}{\sqrt{2} |\eps_K|} (\xi_2 - \xi_0) \,.
\label{eq:e'/e:xi0}
\end{align}
Recently, RBC-UKQCD reported updated results for $\xi_2$ \cite{
Blum:2015ywa}.
The results for $\xi_0$ from the indirect method are given in 
Table \ref{tab:in-LD}.

Recently, RBC-UKQCD has reported new lattice QCD results for $\Im
A_0$ \cite{ Bai:2015nea}.
Combining their results with the experimental value of $\Re A_0$, we
can determine $\xi_0$ directly from the lattice input $\Im A_0$.
This is the direct method.
In Ref.~\cite{ Bai:2015nea}, RBC-UKQCD has also reported the S-wave
$\pi-\pi$ scattering phase shift with isospin $I=0$: $\delta_0 =
23.8(49)(12)$.
This value has $3.0\sigma$ tension with the conventional
determination of $\delta_0$ in Refs.~\cite{ GarciaMartin:2011cn}
(KPY-2011) and \cite{ Colangelo:2001df, Colangelo2016:MITP} (CGL-2001).
They used a singly subtracted Roy-like equation (KPY-2011) or
a doubly subtracted Roy equation (CGL-2001) to do the interpolation
around $\sqrt{s} = m_K$ (kaon mass).
The values for $\delta_0$ are summarized in Table \ref{tab:d0}.
In Fig.~\ref{fig:d0-KPY-2011}, we show the fitting results of KPY-2011.
Their fits to the experimental data work well from the $\pi-\pi$
threshold to $\sqrt{s} = 800 \MeV$, and 
are highly consistent with CGL-2001
in the interpolating region around $\sqrt{s} = m_K \approx 500 \MeV$.
%

%
%
\begin{figure}[htbp]
  \begin{minipage}[b]{0.50\linewidth}
    \begin{center}
      \includegraphics[width=1.0\textwidth]{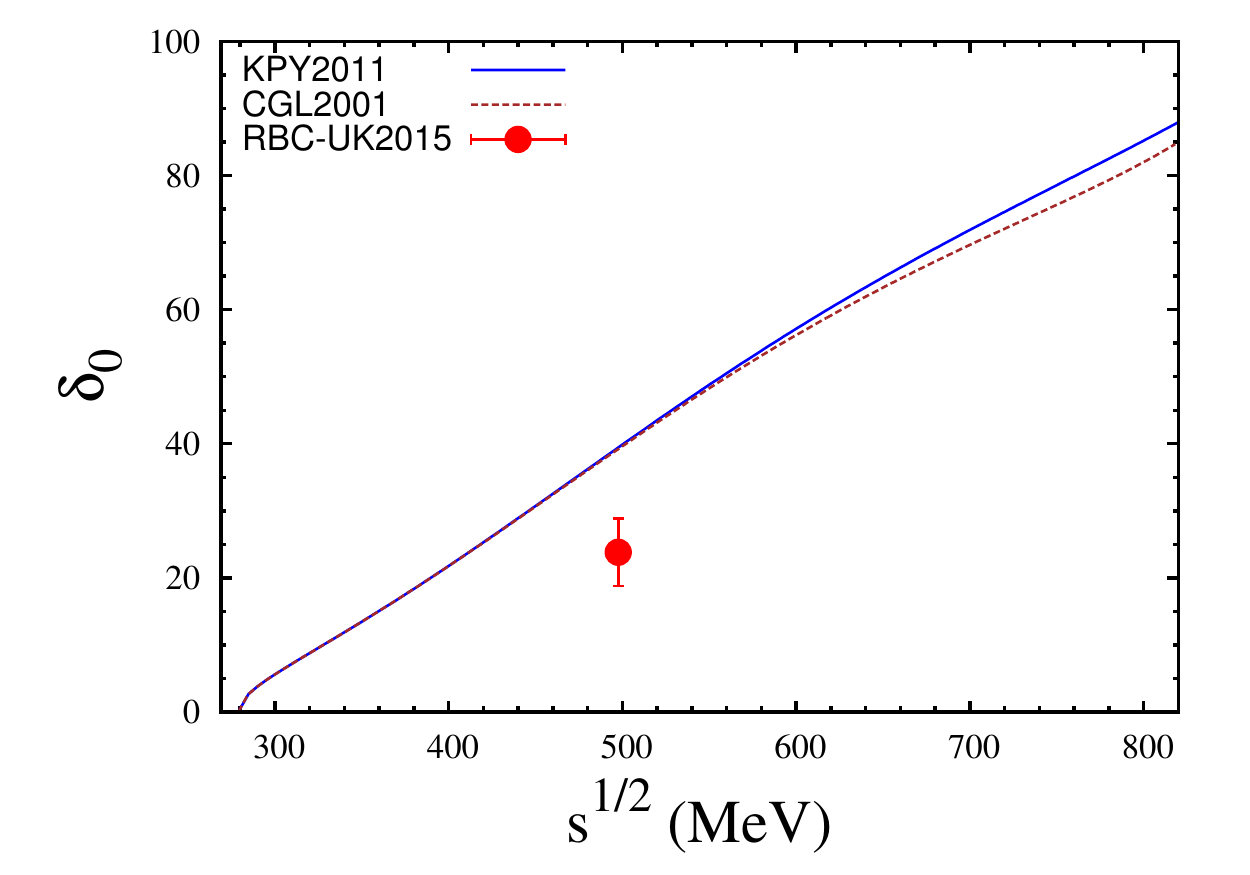}
    \end{center}
    \captionof{figure}{Comparison of $\delta_0$.}
    \label{fig:d0-rbc}
  \end{minipage}
  \hfill
  \begin{minipage}[b]{0.50\linewidth}
    \begin{center}
      \includegraphics[width=1.0\textwidth]{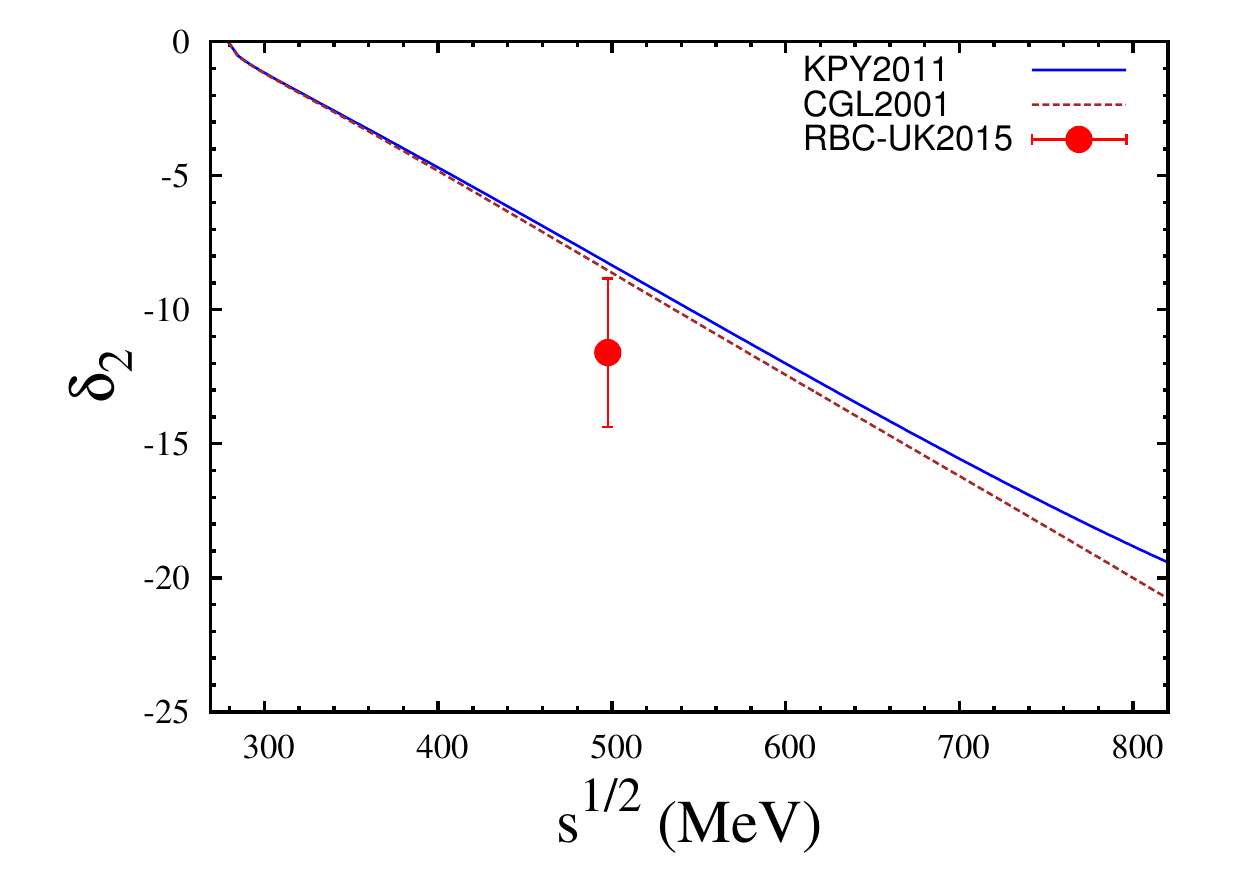}
    \end{center}
    \captionof{figure}{Comparison of $\delta_2$.}
    \label{fig:d2-rbc}
  \end{minipage}
\end{figure}
In Fig.~\ref{fig:d0-rbc}, we present the results of RBC-UKQCD together 
with the fitting results of KPY-2011 and CGL-2001.
There is essentially no difference between KPY-2011 and CGL-2001 in
the region near $\sqrt{s} = m_K \approx 500\MeV$.
Here, we observe the $3.0\sigma$ gap between RBC-UKQCD and KPY-2011.
In contrast, for $\delta_2$ (S-wave, I=2), there is no difference
between RBC-UKQCD and KPY-2011 within statistical uncertainty, 
as one can see in Fig.~\ref{fig:d2-rbc}.

Therefore, we conclude that the results of the indirect method are
more reliable than those of the direct method for $\xi_0$, since the
direct calculation of $\Im A_0$ by RBC-UKQCD might have unresolved
issues.
Hence, we use the indirect method to determine $\xi_0$ in this paper.

Regarding $\xi_\text{LD}$, the long distance effect in the dispersive
part, the theoretical master formula in the continuum is given in
Ref.~\cite{ Bailey:2015tba}.
A theoretical framework for calculating it on the lattice is well
established in Ref.~\cite{ Christ2012:PhysRevD.88.014508}.
There has been an on-going attempt to calculate it on the lattice
\cite{ Christ:2014qwa}.
However, this attempt \cite{ Bai2016:Latt}, at present, is not mature
and belongs to the category of exploratory study rather than to that
of precision measurement.
Hence, we use a rough estimate of $\xi_\text{LD}$ given in
Ref.~\cite{ Christ:2014qwa} in this paper.  
It is summarized in Table \ref{tab:in-LD}.

\subsection{$\BK$}

In Table \ref{tab:in-BK}, we present results for $\BK$ calculated
using lattice QCD tools with $N_f=2+1$ flavors.
Here, FLAG-2016 represents the global average of the results of
BMW-2011 \cite{ Durr2011:PhysLettB.705.477}, Laiho-2011 \cite{
  Laiho:2011np}, RBC-UK-2016 \cite{ Blum:2014tka}, and SWME-2016
\cite{ Jang:2015sla}, which is summarized in Ref.~\cite{
  Aoki:2016frl}.
SWME-2014 represents the $\BK$ result reported in Ref.~\cite{
Bae2014:prd.89.074504}.
RBC-UK-2016 represents that reported in Ref.~\cite{ Blum:2014tka}.

The results of SWME-2016 are obtained using fitting based on staggered
chiral perturbation theory (SChPT) in the infinite volume limit, and
those of SWME-2014 are obtained using fitting based on SChPT with finite
volume corrections incorporated at the NLO level.
Here we use the FLAG-2016 result for $\BK$.
\begin{table}[htbp]
  \begin{minipage}[b]{0.40\textwidth}
    \caption{$\BK$.}
    \label{tab:in-BK}
    \begin{center}
    \small
    \begin{tabular}{lll}
      \br
      Collaboration & Value & Ref.
      \\ \mr
      FLAG-2016   & $0.7625(97)$      & \cite{Aoki:2016frl}
      \\
      SWME-2014   & $0.7379(47)(365)$ & \cite{Bae2014:prd.89.074504}
      \\
      RBC-UK-2016 & $0.7499(24)(150)$ & \cite{Blum:2014tka}
      \\ \br
    \end{tabular}
    \end{center}
  \end{minipage}
  \hfill
  \begin{minipage}[b]{0.58\textwidth}
    \caption{Wolfenstein parameters.}
    \label{tab:in-CKM}
    \begin{center}
    \small
    \begin{tabular}{llll}
      \br
      & CKMfitter & UTfit & AOF
      \\ \mr
      $\lambda$
      & $0.22548(68)$/\cite{Charles:2015gya}
      & $0.22497(69)$/\cite{UTfit2016:web}
      & $0.2253(8)$/\cite{Agashe2014:ChinPhysC.38.090001}
      \\
      $\bar{\rho}$
      & $0.145(13)$/\cite{Charles:2015gya}
      & $0.153(13)$/\cite{UTfit2016:web}
      & $0.139(29)$/\cite{UTfit2014PostMoriondSM:web}
      \\
      $\bar{\eta}$
      & $0.343(12)$/\cite{Charles:2015gya}
      & $0.343(11)$/\cite{UTfit2016:web}
      & $0.337(16)$/\cite{UTfit2014PostMoriondSM:web}
      \\ \br
      \end{tabular}
    \end{center}
  \end{minipage}
\end{table}

\subsection{Other input parameters}

For the Wolfenstein parameters $\lambda$, $\bar{\rho}$, and
$\bar{\eta}$, both CKMfitter and UTfit updated their results in
Refs.~\cite{ Charles:2015gya, UTfit2016:web}, but the angle-only-fit
has not been updated since 2015.
The global unitarity triangle (UT) fits of both CKMfitter and UTfit
use $\epsK$ and $|V_{cb}|$ as input parameters to determine the apex
$\bar{\rho}$ and $\bar{\eta}$.
Hence, using them to evaluate $\epsK$ introduces unwanted
correlations through $\epsK$ and $|V_{cb}|$.
In contrast, the angle-only-fit (AOF) results are independent of $\epsK$
and $|V_{cb}|$.
Hence, we use the AOF results in this paper.

For the QCD corrections $\eta_{cc}$, $\eta_{ct}$, and
$\eta_{tt}$, we use the same values as in Ref.~\cite{ Bailey:2015tba},
which are given in Table \ref{tab:in-qcd}.
In particular, we use the SWME value of $\eta_{cc}$ reported in
Ref.~\cite{ Bailey:2015tba} instead of that in Ref.~\cite{
  Brod2011:PhysRevLett.108.121801}.
This issue is well explained in Ref.~\cite{ Bailey:2015tba}.
One of the reasons is that the size of the NNLO correction 
is already a conservative estimate for the truncation error of
the NNNLO level in perturbation theory.
Another reason is that the SWME result is highly consistent with
that of Ref.~\cite{ Buras2013:EurPhysJC.73.2560}.
\begin{table}[htbp]
  \begin{minipage}[c]{0.4\textwidth}
    \caption{QCD corrections.}
    \label{tab:in-qcd}
    \begin{center}
    \begin{tabular}{lll}
      \br
      Input & Value & Ref. 
      \\ \mr
      $\eta_{cc}$ & $1.72(27)$
      & { \protect\cite{Bailey:2015tba} } \\
      $\eta_{tt}$ & $0.5765(65)$
      & { \protect\cite{Buras2008:PhysRevD.78.033005} } \\
      $\eta_{ct}$ & $0.496(47)$
      & { \protect\cite{Brod2010:prd.82.094026} }
      \\ \br
    \end{tabular}
    \end{center}
  \end{minipage}
  \hfill
  \begin{minipage}[c]{0.6\textwidth}
    \caption{Other input parameters.}
    \label{tab:in-other}
    \begin{center}
    \begin{tabular}{lll}
      \br
      Input & Value & Ref. 
      \\ \mr
      $G_{F}$
      & $1.1663787(6) \times 10^{-5}$ GeV$^{-2}$
      &\cite{Agashe2014:ChinPhysC.38.090001} \\
      $M_{W}$
      & $80.385(15)$ GeV
      &\cite{Agashe2014:ChinPhysC.38.090001} \\
      $m_{c}(m_{c})$
      & $1.2733(76)$ GeV
      &\cite{Chakraborty:2014aca} \\
      $m_{t}(m_{t})$
      & $163.3(2.7)$ GeV
      &\cite{Alekhin2012:plb.716.214} \\
      $\theta$
      & $43.52(5)^{\circ}$
      &\cite{Agashe2014:ChinPhysC.38.090001} \\
      $m_{K^{0}}$
      & $497.614(24)$ MeV
      &\cite{Agashe2014:ChinPhysC.38.090001} \\
      $\Delta M_{K}$
      & $3.484(6) \times 10^{-12}$ MeV
      &\cite{Agashe2014:ChinPhysC.38.090001} \\
      $F_K$
      & $156.2(7)$ MeV
      &\cite{Agashe2014:ChinPhysC.38.090001}
      \\ \br
    \end{tabular}
    \end{center}
  \end{minipage}
\end{table}

In Table \ref{tab:in-other}, we summarize other input parameters.
They are the same as those in Ref.~\cite{ Bailey:2015tba} except
for the charm quark mass $m_c(m_c)$.
For the charm quark mass, we use the HPQCD result reported in
Ref.~\cite{ Chakraborty:2014aca}.

\section{Current status of $\epsK$}
\begin{figure}[htbp]
  \begin{minipage}[b]{0.49\linewidth}
    \begin{center}
      \includegraphics[width=1.0\textwidth]{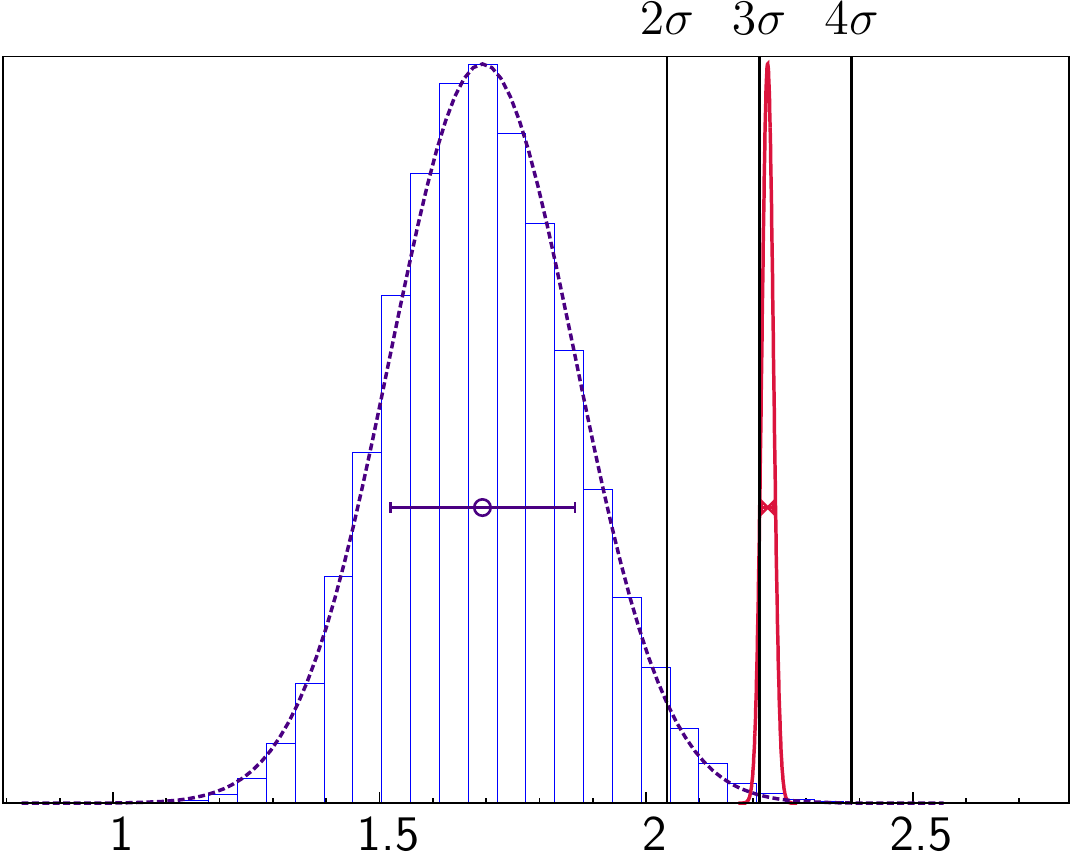}
    \end{center}
    \captionof{figure}{$\epsK$ with exclusive $V_{cb}$.}
    \label{fig:eps-ex}
  \end{minipage}
  \hfill
  \begin{minipage}[b]{0.49\linewidth}
    \begin{center}
      \includegraphics[width=1.0\textwidth]{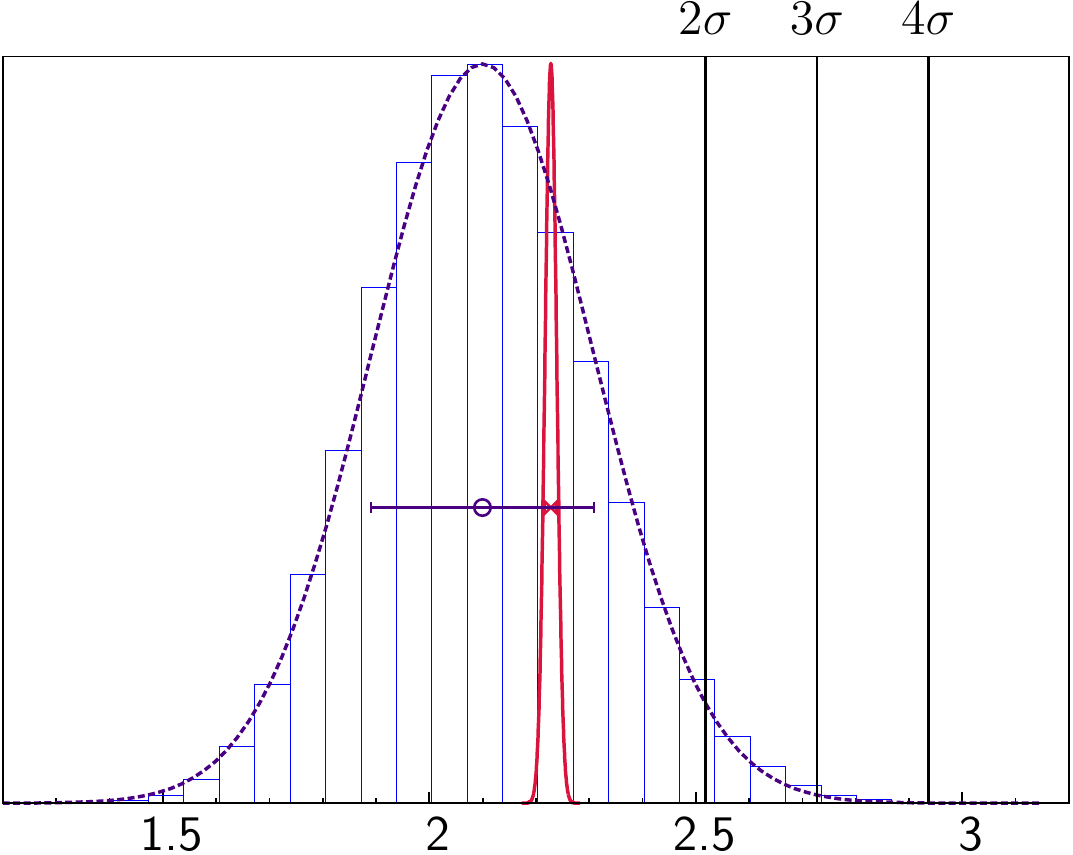}
    \end{center}
    \captionof{figure}{$\epsK$ with inclusive $V_{cb}$.}
    \label{fig:eps-in}
  \end{minipage}
\end{figure}

In Fig.~\ref{fig:eps-ex}, we show the results for $\epsK$ evaluated 
directly from the standard model with the lattice QCD inputs described
in the previous section.
Here, the blue curve represents the theoretical evaluation of
$\epsK$ with the FLAG-2016 $\BK$, AOF for the Wolfenstein parameters,
and exclusive $|V_{cb}|$ that corresponds to ex-combined in Table
\ref{tab:Vcb}.
The red curve represents the experimental value of $\epsK$.
In Fig.~\ref{fig:eps-in}, the blue curve represents the same as that
in Fig.~\ref{fig:eps-ex} except for using the inclusive $|V_{cb}|$
in Table \ref{tab:Vcb}.
Our preliminary results are, in units of $1.0 \times 10^{-3}$,
\begin{align}
  |\epsK| &= 1.69 \pm 0.17 && \text{for exclusive $V_{cb}$ (lattice QCD)}
  \\
  |\epsK| &= 2.10 \pm 0.21 && \text{for inclusive $V_{cb}$ (QCD sum rules)}
  \\
  |\epsK| &= 2.228 \pm 0.011 && \text{(experimental value)}
\end{align}
We find that there is $3.2\sigma$ tension in the exclusive $V_{cb}$
channel (lattice QCD), and no tension in the inclusive $V_{cb}$
channel (heavy quark expansion; QCD sum rules).
%
%
%
%


\ack 
We thank R.~Van de Water for helpful discussion on $V_{cb}$.
The research of W.~Lee is supported by the Creative Research
Initiatives Program (No.~20160004939) of the NRF grant funded by the
Korean government (MEST).
W.~Lee would like to acknowledge the support from the KISTI
supercomputing center through the strategic support program for the
supercomputing application research (No.~KSC-2014-G3-003).
The computations were carried out in part on the DAVID GPU clusters at
Seoul National University.

\section*{References}
\bibliography{ref}

\end{document}